# High-temperature and high-pressure apparatus aiming for synthesis of solid metallic hydrogen

Yasushi Kawashima
Department of Precision Engineering, School of Engineering, Tokai University, Hiratsuka, Kanagawa 259-1292, Japan.

E-mail: kawasima@keyaki.cc.u-tokai.ac.jp

**It was predicted that solid metallic hydrogen can be obtained if solid molecular hydrogen is pressured to high pressure at low temperature about 80 years ago. Furthermore, the solid metallic hydrogen was theoretically predicted to show superconductivity at room temperature. In addition, surprising prediction was made that the solid metallic hydrogen is a metastable metal with a potential barrier of 1 eV. This prediction implies that the solid metallic hydrogen remains in the metallic solid phase state even after it is released from high pressure. Shock compression synthesized liquid metallic hydrogen so far. However, to obtain metallic hydrogen under ambient pressure at room temperature, it is necessary to synthesize solid metallic hydrogen. It is theoretically predicted that an ultrahigh pressure of ~500 GPa is required to produce solid metallic hydrogen. This pressure is close to the limit of the high pressure that can be generated by the diamond anvil cell (DAC). Therefore, it is difficult to synthesize metallic hydrogen by using DAC. Shock compression methods can generate a sufficiently high pressure to synthesize metallic hydrogen. But it is impossible to synthesize solid metallic hydrogen by these methods because these accompany generation of high temperature. There is currently no definitive way to synthesize solid metallic hydrogen.**

    **Here we report a new dynamic high-pressure apparatus which aims for synthesis of solid metallic hydrogen. We show that by using this apparatus, it is possible to generate a high pressure of 1 TPa, the pressure maintaining time is $10^3$ to $10^6$ times longer than the conventional dynamic compression method. Furthermore, we show that this apparatus is not accompanied by the generation of high temperature unlike the conventional dynamic compression method and then it is possible that solid metallic hydrogen might be synthesized by this apparatus.**

# 1. Introduction

In 1935, Wigner and Huntington anticipated that if the solid molecular hydrogen, which is an insulator, was pressurized to a pressure higher than 25 GPa at very low temperature, this solid molecule would dissociate at a density of 0.62 molH/cm$^3$ and it would undergo a dissociative transition to an atomic solid with a half-filled conduction band so that it would become metallic [1]. Subsequently, Ashcroht suggested that this putative metal hydrogen is a high temperature superconductor [2], and that it may be a room temperature superconductor [3]. In addition, important theoretical predictions were indicated by Brovmen, Kagan, and Kholas [4]. That is that solid metal hydrogen is a metastable metal with a potential barrier of 1 eV [4]. This prediction means that as with the diamond which is the metastable phase of carbon, once if the solid metallic hydrogen is synthesized under high pressure, it remains a metallic phase even after it has been released from the pressure and it can be obtained at room temperature under atmosphere. However, no solid metallic hydrogen has been synthesized so far, none of the above theoretical predictions have been confirmed.

If solid metal hydrogen is synthesized, it is not only possible to obtain room temperature superconductors, but it has also been pointed out that it may revolutionize rocket engineering as a powerful propellant [5].

Hydrogen does not metallize up to 340 GPa at room temperature [6, 7]. On the other hand, it was observed by the diamond anvil high pressure apparatus that solid hydrogen becomes completely opaque under the conditions of 100 K and 320 GPa [8]. According to the recent quantum Monte Carlo method and density functional theory, it is predicted that a pressure of 400-500 GPa is required for metallization of hydrogen molecules [9-11]. It is also predicted that solid hydrogen molecules will not dissociate and metallize until it exceeds the pressure of ~500 GPa above the high-pressure generation limit of conventional diamond anvil high pressure apparatus [12]. The diamond anvil high pressure apparatus is thought to be capable of generating static high pressure of 250 to 400 GPa, but it is not easy to generate even such pressure range. Even if metallic hydrogen is synthesized by the diamond anvil high pressure apparatus, the volume of the high-pressure region in the apparatus is very small. Although its volume depends on pressure, it is 10$^{-6}$ to 10$^{-1}$ mm$^3$. On the other hand, the dynamic compression method using shock waves can easily achieve a pressure of 100 to 1000 GPa, but the dynamic pressure duration is very short. In addition, dynamic high-pressure generation usually accompanies generation of high temperature of several thousand degrees [13, 14]. For example, the shock temperatures of iron at high pressures of 70 GPa and 300 GPa are 1300 K and 10,000 K, respectively [15]. Even in the case of high strength alumina with a relatively low compression ratio, a sudden temperature rises due to the shock stress of 100 GPa estimated by calculation is ~ 1000 K [16]. In 1996 the metallic phase of hydrogen was found with a density of 0.64 mol H/cm$^3$, a pressure of 140 GPa, a temperature of 3000 K, by dynamic compression with a duration of ~ 100 ns [17]. Since the melting point of hydrogen at 140 GPa is ~900 K, the synthesized metallic hydrogen is liquid metallic hydrogen. Liquid metallic hydrogen cannot maintain the metal phase after removing pressure. The engineering significance lies in the synthesis of solid metal hydrogen, which is expected to maintain the metallic state after removing pressure. Currently, we have no choice but to try solid-state metal hydrogen synthesis using only a diamond anvil high pressure apparatus. However, the lowest predicted pressure for synthesizing solid metallic hydrogen is close to the highest limit pressure that can be generated by the diamond anvil high pressure device [12].

Silvera et. al. attempted to synthesize solid metallic hydrogen by careful experiments so that the diamond anvil would not be broken, using a diamond anvil high pressure apparatus [18]. As a result, they showed that hydrogen in the pressure chamber reflected light at 495 GPa [18]. Based on this result, they argued that the transition of solid molecular hydrogen to the solid metallic hydrogen occurred. However, a researcher pointed out that the reflection of light observed may be due to the metallization of the alumina film which covers the diamond anvil pressure surface to prevent brittleness of diamond by hydrogen and he denied this result [19]. Furthermore, it has been reported that metallic hydrogen, which was considered to have been synthesized, disappeared. After all, it was not ascertained that after the pressure was removed, the solid metal hydrogen was metastable in atmospheric pressure.

It is technically difficult to synthesize metallic hydrogen using diamond anvil high pressure apparatus. Although the dynamic compression method can generate sufficiently high pressure to synthesize metallic hydrogen, the generation of high pressure involves the generation of high temperature. Therefore, this method is not suitable for the synthesis of solid metallic hydrogen. Then, in this report, we introduce a high temperature and high-pressure apparatus which can be aimed for synthesis of solid metallic hydrogen. In this apparatus, the pressure source material is constrained by a transparent material and a part of the pressure source is vaporized by pulsed laser irradiation. High pressure is generated by constraining explosion force produced by the decomposition of this pressure source [20-22]. We show that this apparatus can generate a high pressure of 1 TPa, it can maintain high pressure $10^3$-$10^6$ times longer than the conventional dynamic compression method and it is not accompanied by the generation of high temperature when generating high pressure.

## 2. Apparatus
### 2.1 Principle

Bridgman developed a method for generating high pressure by mechanical compression [23]. Although this method sacrifices the hydrostatic pressure, it became possible to obtain high pressure several orders of magnitude higher than the high pressure obtained so far. However, since there is a sliding part in this method, most of the external forces applied do not reach the pressurized sample because of the friction and the pressurization of the gasket. Even in the case of the dynamic high-pressure generation method, for the same reason, most of the applied force is not transmitted to the sample to be pressurized. Instead of applying force from the outside, a method was developed to generate high pressure by generating force inside. In this method, by concentrating electromagnetic energy on an object surrounded perfectly without gaps by a material, heat energy is applied to the object to expand it, thereby generating high pressure by constraining the expansion of the object [24-27]. It was shown that high pressure capable of synthesizing diamond can be generated by combination of materials constituting the device [26]. In this method, the above-mentioned problem in the mechanical compression method by Bridgman is solved.

If some or most of the interatomic bonds in an object that is completely constrained around is decomposed and the part or most of the object explodes, the rest of the object should shrink its volume significantly. The shrinkage rate is several orders of magnitude larger than in the case of constraining thermal expansion. If energy is externally applied to decompose the interior into discrete atoms, an explosive force is generated inside and ultrahigh pressure can be generated with very high efficiency. In this research, a method of generating ultra-high pressure by constraining volume expansion due to decomposition of interatomic bonds is proposed [20-22].

### 2.2 Method

The pressure generating source is surrounded by a high-strength material which is partly optically transparent. Then, the pressure source is decomposed by irradiating a part of the pressure source through the light-transmissive portion with pulsed laser light. Constraining expansion due to decomposition of the pressure source generates high pressure. As an example, a method for generating the high pressure is shown in Fig. 1. As shown in Fig. 1, two opposed light-transmissive anvils are used to constrain the pressure generating source. A part of the pressure generating source is decomposed by irradiating the pulse laser to it via one anvil or both anvils. High pressure generation can be efficiently performed by using a material having high light absorption property and strong interatomic or intermolecular bonding force as a pressure generating source.

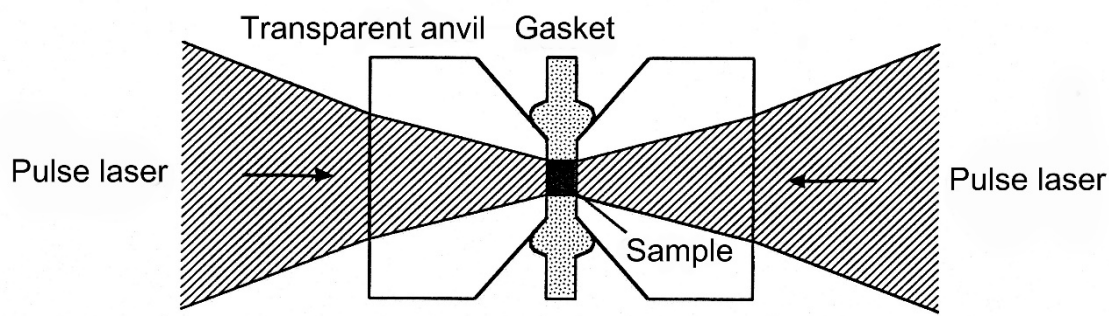

Figure 1 **Method of constraining expansion of substance and method of energy input.**

**2.3 Configuration of high-pressure apparatus**

In the proposed ultrahigh pressure apparatus, a single pulse laser beam having an arbitrary pulse width of 0.1 ms to 2 ms emitted from a pulse YAG laser apparatus is irradiated to a pressure source pressurized by a diamond anvil high pressure cell (DAC) or a sapphire anvil high pressure cell (SAC), and an ultrahigh temperature region is thereby formed in a part thereof. Ultrahigh pressure is generated by constraining the expansion force caused by vaporization of pressure source in the ultra-high temperature region. The apparatus consists of a DAC or a SAC, a pulsed YAG laser apparatus (YP 500 A, FANUC Co., Ltd.), an optical system for laser light irradiation microscope, and an optical system for ruby fluorescence spectrum or radiation spectrum measurement [20].

A schematic diagram of the ultrahigh pressure apparatus and the temperature and pressure measurement system is shown in Fig 2. In Fig. 2, when the diamond anvil cell (DAC) or the sapphire anvil cell (SAC) high pressure apparatus is viewed as the center, the right side is the optical system for ruby fluorescence spectrum measurement or the optical system for radiation spectrum measurement, the left side is the laser light irradiation microscope optical system. In this laser irradiation microscope, the laser irradiation spot can be set at an arbitrary position within the monitor viewing field, and the laser spot diameter can be changed from 50 μm to 1 mm. Adjustment of the laser beam irradiation position and the irradiation diameter is performed by using the He-Ne laser guide light emitted from the pulsed YAG laser apparatus. The maximum peak power and maximum peak energy of the pulse that the YAG laser apparatus can emit are 35 kW and 70 J respectively, the wavelength is 1.06 μm, the beam quality is less than 15 mm-mrad, the pulse width can be changed from 0.2 ms to 2 ms every 0.1 ms, and the pulse waveform can be controlled in units of 0.1 ms. The DAC or SAC is installed on the XYZ stage, and the position of the anvil cell can be changed with respect to the objective lens. Thus, it is possible to focus the pulsed YAG laser light on an arbitrary position of the sample.

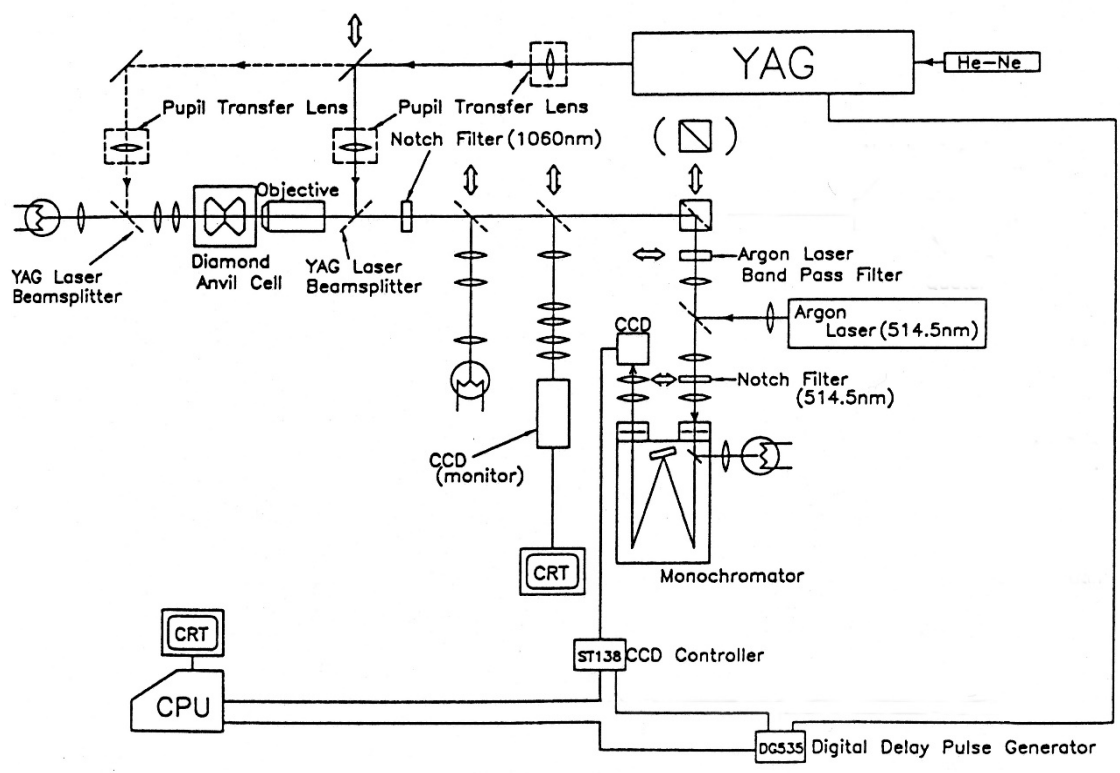

Figure 2 **Schematic diagram of the high-pressure apparatus and the temperature and pressure measurement system.**

The laser light emitted from the YAG laser apparatus is sent to the YAG laser splitter by pupil shift system and is reflected thereby and irradiated to the sample in the DAC or SAC through the objective lens of the laser light irradiation microscope. Argon laser light (wavelength: 514.5 nm) for ruby fluorescence spectrum measurement passes through the objective lens of the optical system for ruby fluorescence spectrum measurement and is irradiated to an arbitrary ruby grain located in the DAC or SAC. The ruby fluorescence from the ruby grain excited by the argon laser light passes through the same optical path as the irradiated argon laser in the opposite direction and reaches the notch filter for the wavelength 514.5 nm. Light passing through the notch filter is dispersed by the spectroscope and reaches the CCD detector. The apparatus is equipped with epi-illumination for observation of the sample. This apparatus is designed so that an arbitrary position in the sample can be selected as the ruby fluorescence measurement point while observing the sample with the CCD monitor. The CCD is controlled by the time-resolved spectroscopic analysis mode (kinetics mode) to record the time change of the ruby fluorescence spectrum or the radiation spectrum at high speed. The YAG laser and the CCD are controlled by a computer via a Digital Pulse & Delay Generator (STANFORD RESEARCH SYSTEMS CORPORATION, DG 535). In this way, the timings of the respective devices are set so that the emission of the pulsed laser light from the YAG laser apparatus and the time-resolved measurement by the CCD are simultaneously started.

In the ruby fluorescence spectrum measurement system, argon laser light for ruby fluorescence excitation was irradiated to an arbitrary point in the region of φ 50 μm in the diamond anvil cell or sapphire anvil cell pressure chamber, and the emission spectrum of ruby from this position was measured with a spectrometer (HR 320, 150 gr/mm, JOBIN YVON, Ltd) and a cooled CCD detector (TE/CCD 1024 E, Princeton Instruments). This cooled CCD detector is controlled in a time-resolved spectroscopic analysis mode (kinetics mode) by using a detector controller (ST-130, Princeton Instruments). This measurement mode makes it possible to measure fast time changes of various spectra. As a probe laser, an argon laser (5145 Å) s used. The minimum narrowing diameter of this laser light is about 5 μm. Using this measurement system, ruby grains located in a diamond or sapphire anvil high pressure chamber can be selected and irradiated with argon laser light to measure the ruby fluorescence spectrum. The maximum time division of the time-resolved measurement in the kinetics mode of the CCD detector is 6 μs as the one-pixel sweeping speed, and the time division data can measure 256 data in the pixel conversion.

As a sapphire anvil cell device, a clamp type device was used [28]. A cross-sectional view of this device is shown in Fig. 3. In this device, the pressure is generated by tightening the upper and lower anvils with three bolts. As shown in Fig. 3, the upper anvil is adhered onto a cemented carbide pedestal which is attached to a spherical seat. The cemented carbide pedestal to which the lower anvil is adhered can move in parallel and its position can be adjusted so that the center of the upper anvil pressure plane and the center of the lower anvil pressure plane are aligned. Furthermore, the parallelism of the anvil pressure plane can be adjusted by the screws attached to the spherical seat. Also, to prevent damage to the anvil due to excessive tightening of the bolt, it can be prevented with three screws. The body is a heat-treated copper beryllium alloy.

The anvil pressure surface of the sapphire anvil used was polished to be an alumina (111) surface. The anvil pressure surface diameter, bottom surface diameter, and height are 1 mm, 5 mm, and 3 mm, respectively. The size and shape of the sapphire anvil are shown in Fig. 4.

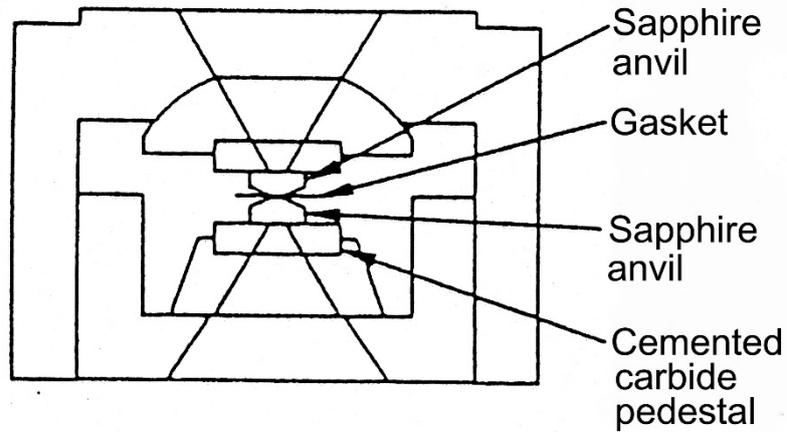

Figure 3 **Sapphire anvil high pressure cell.**

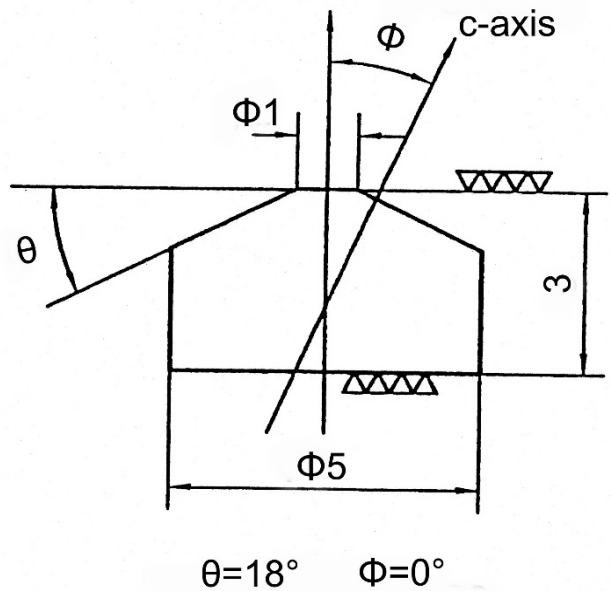

Figure 4 **Shape and dimensions of sapphire anvil.**

## 3. High pressure experiment and ruby fluorescence measurement
### 3.1 Measurement methods
#### 3.1.1 Time resolved measurement of radiation spectrum

A schematic diagram of a sapphire anvil high pressure apparatus and a high temperature measurement apparatus is shown in Fig. 5. As shown in Fig. 5, the pulsed YAG laser light passes through the laser irradiation optical system, passes through a hole of 1 mm in diameter of the mirror 1, and is irradiated to the sample chamber of the sapphire anvil. The laser spot diameter in the sample is adjusted by the laser irradiation optical system. Radiant light from the high temperature part of the sample is collected by the objective lens for laser irradiation and is sent to the spectroscope and the CCD detector as parallel light by using the mirror 1, the mirror 2, the mirror 3, and the mirror 4 (see Fig. 5). As shown in the inset of Fig. 5, the pressure generating source that is irradiated with the laser light is a glassy carbon powder pressurized by a sapphire anvil, and the sample configuration in the SAC is the same as that in the ruby fluorescence measurement experiment. The glassy carbon powder layer was irradiated with a pulse laser light having a pulse width of 0.4 ms and $2.1 \times 10^{7}$-W/cm$^2$.

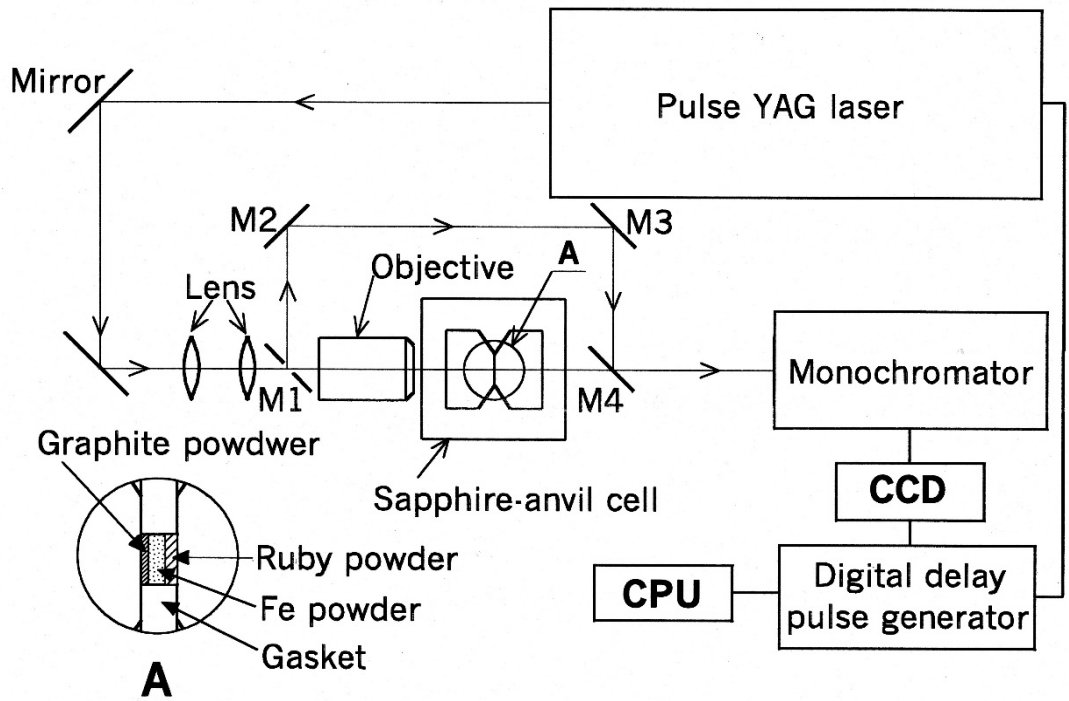

Figure 5 **Schematic diagram of the apparatus for measuring the temperature of the sample surface irradiated with pulsed YAG laser.**

Since the spectrum to be measured is largely affected by the sensitivity difference of the CCD depending on the wavelength and the characteristics of the optical system, it is necessary to correct the spectrum taking these factors into consideration. Therefore, the radiation spectrum from the spectral irradiance standard lamp from which spectral irradiance was obtained was measured. The following equation (1) gives the true radiation spectrum.

[*True radiation data of the sample*] = [*Validation radiation data obtained for the spectral irradiance standard lamp*] x [*Radiation data measured for the sample using the optical system*] / [*Radiation data measured for the spectral irradiance standard lamp by using the optical system*]      (1)

The spectrum from the high-temperature part instantaneously generated by pulsed laser irradiation was measured by using CCD detector in kinetic mode. In this mode, spectra are stored every 6 μs in CCD pixel lines (1024 pixels).

### 3.1.2 Time-resolved measurement of fluorescence spectrum from ruby grains

Figure 6 shows a method of constraining the expansion of sample with two transparent anvils, a method of irradiating the sample with a pulsed laser light through one anvil, and a method of measuring fluorescence emission from ruby grains located on the other anvil pressure surface. Although the compressive strength of sapphire is considerably lower than that of diamond, sapphire was used for constraining the expansion of the sample for the following reasons. It is because sapphire does not emit high intensity fluorescence unlike diamonds [29, 30] and a large synthetic single crystal sapphire is available. Glassy carbon was chosen as a pressure source (an object that generates its expansion force by breaking its interatomic bonds) because of its large light absorption coefficient, strong covalent bond, and very small thermal diffusion coefficient. A carbon powder layer was formed on the pulsed YAG laser irradiation anvil face and a ruby powder layer was formed on the opposite anvil face (Fig. 6).

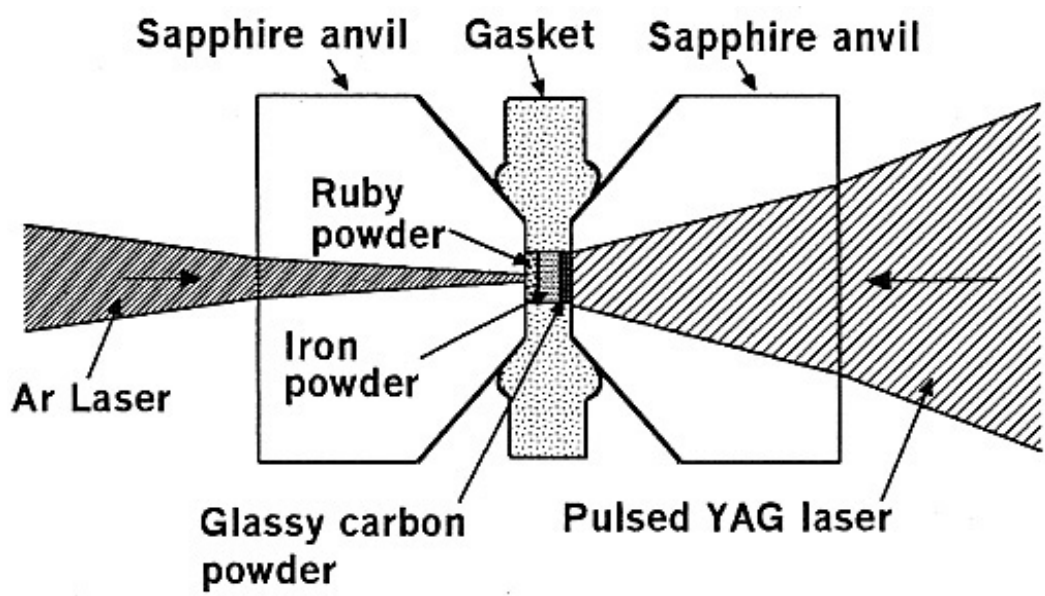

Figure 6 **Ultrahigh pressure generation method and ruby fluorescence measurement method.**

An iron powder layer was sandwiched between glassy carbon and ruby layers as a pressure transmission medium. Ruby, iron, and carbon powders were packed in 0.15 mm diameter holes drilled in a copper beryllium flat plate with a thickness of 0.2 mm and they were pressurized at 5 GPa by the two pressure faces of the sapphire anvils. A screw clamp type anvil cell was used to pressurize the sample. To further pack the powder samples in the gasket hole, the glassy carbon layer was irradiated with weak pulsed laser, and after the irradiation, the three screws of the sapphire anvil cell were slightly tightened. The size of the sample chamber was about 0.15 mm in diameter and about 0.15 mm in height. Ruby fluorescence was excited with a 514.5 nm line of an argon laser. After ruby fluorescence passed through a holographic notch filter, it was dispersed by a monochromator and detected by a cooled CCD camera with 1024 x 256 pixels [20]. The ruby fluorescence spectrum was continuously recorded every 30 μs on the CCD pixel line.

## 3.2 Results and discussion
### 3.2.1 Measurement of radiation spectrum from carbon layer and measurement of generated temperature

The radiant energy emitted from the unit area of the blackbody with the temperature $T$ in unit time is approximately represented by Wien's equation as follows.

$$W(\lambda, T) = c_1 \lambda^{-5} \exp(-\frac{c_2}{\lambda T}) \qquad (2)$$

Here, $W(\lambda, T)$ is the energy component of the wavelength $\lambda$ of the energy radiated from the object at the temperature $T$, $c_1 = 1.191 \times 10^{-12}$ W·cm$^2$, $c_2 = 1.438$ cm·K. The following equation (3) gives the radiated energy from a real object that is not a black body.

$$W'(\lambda, T) = \tau c_1 \lambda^{-5} \varepsilon(\lambda, T) \exp(-c_2/\lambda T) \qquad (3)$$

In the above equation (3), $\tau$ is the transmittance of an absorbing object such as water vapor, glass, etc., which exists between the measured object and the temperature measurement device. $\varepsilon(\lambda, T)$ is the emissivity, which varies depending on the temperature, material, surface condition, etc. of the object and has a value between 0 and 1. Furthermore, the emissivity is not constant for the radiant energy of all wavelengths. In the visible light region, the change of emissivity with wavelength is small. The emissivity gradually decreases as the wavelength becomes longer in the wavelength region longer than the infrared light region. Radiation energies $W_1$ and $W_2$ of two different wavelengths $\lambda_1$ and $\lambda_2$ of radiant light emitted from an object at a temperature $T$ are expressed by the following equations (4) and (5), respectively. When these ratios are taken, the equation (6) is obtained.

$$W_1' = \tau_1 c_1 \lambda_1^{-5} \varepsilon_1(\lambda_1, T)\exp(-c_2/\lambda_1 T) \qquad (4)$$

$$W_2' = \tau_2 c_1 \lambda_2^{-5} \varepsilon_1(\lambda_2, T)\exp(-c_2/\lambda_2 T) \qquad (5)$$

$$R = \frac{W_1'}{W_2'} = \frac{\tau_1}{\tau_2}\left(\frac{\lambda_1}{\lambda_2}\right)^{-5}\frac{\varepsilon_1}{\varepsilon_2}\exp\left\{\frac{c_2}{T}\left(\frac{1}{\lambda_2}-\frac{1}{\lambda_1}\right)\right\} \qquad (6)$$

Even if a light absorbing object exists between the measured object and the temperature measurement device, it can be regarded as $\tau_1 \approx \tau_2$ unless it absorbs light of a specific wavelength. Therefore, since $\tau_1/\tau_2 \approx 1$, the transmittance $\tau$ does not affect the ratio $R$. Even if the emissivity $\varepsilon_1$ and $\varepsilon_2$ are unknown, $\varepsilon_1$ can be regarded as being almost equal to $\varepsilon_2$ (i.e. $\varepsilon_1/\varepsilon_2 \approx 1$) when the wavelengths $\lambda_1$ and $\lambda_2$ are close to each other with respect to the wavelength in the visible light region. Therefore, even if the emissivity is unknown and an unknown light absorber exists between the measured object and the radiation spectrum measurement system, the temperature of the object can be obtained by the intensity ratio of two neighboring wavelengths or the local spectrum in the visible region.

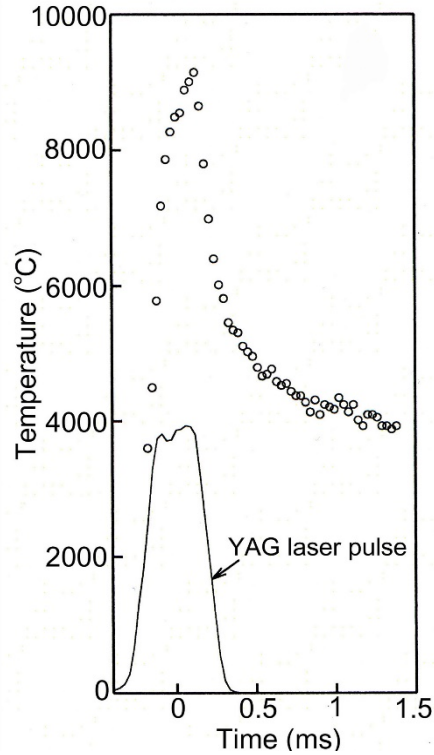

Figure 7 **Change with time of temperature of interface between sapphire anvil pressure surface and pulsed YAG laser irradiated glassy carbon layer**

The radiation spectrum in the range from 650 to 800 nm measured through the sapphire anvil was corrected by using the equation (1) described in 3. 1. 1 with reference to the radiation spectrum from the standard lamp. Further, the radiation spectrum curves fitted to the radiation spectrum data in the range of 650 to 700 nm were obtained by the least square method. The change with time in temperature of the laser irradiated surface in the glassy carbon layer was obtained from these radiation spectrum curves. Figure 7 shows the changes with time in temperature at the boundary between the glassy carbon layer and the sapphire anvil pressure surface after laser irradiation. Figure 7 also shows the change with time of the intensity of the pulsed YAG laser. It is seen from Fig. 7 that the temperature reached the maximum temperature ~9200 °C at 0.174 ms after laser irradiation, and then gradually decreased. The measurement result shows that by pulsed laser irradiation with a pulse width of 0.4 ms and $2.1 \times 10^7$ W/cm$^2$, the temperature of the glassy carbon layer rose to a temperature (~ 4000 K) [31] at which the carbon decomposes.

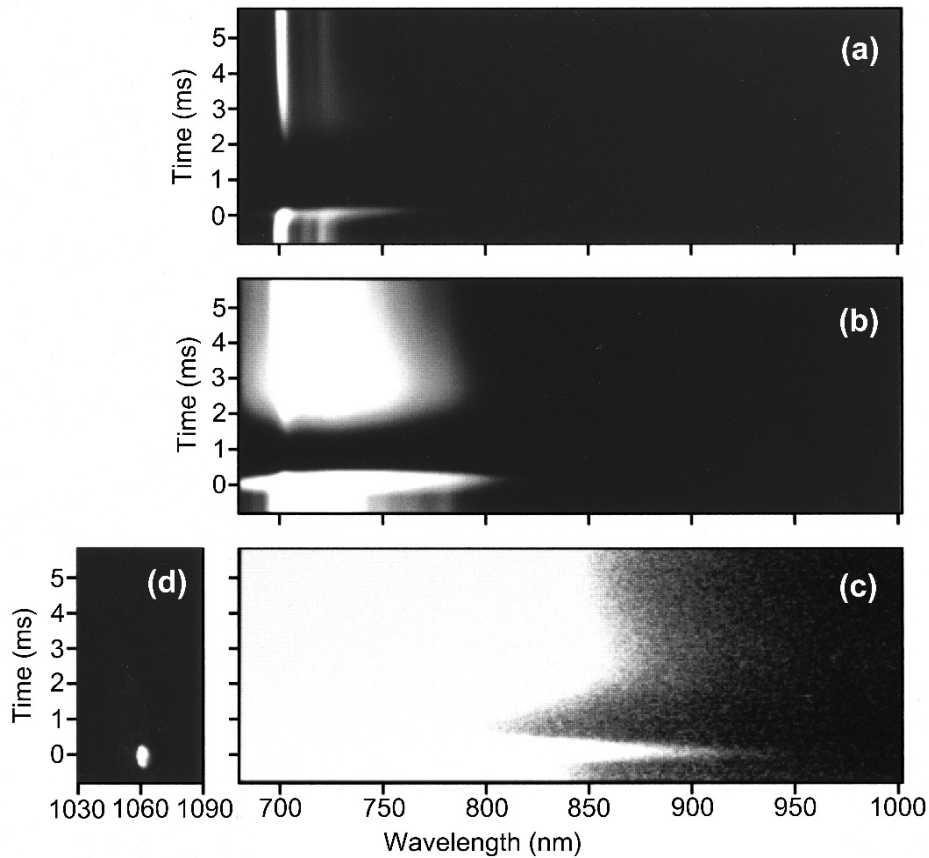

Figure 8 **CCD camera recording of ruby fluorescence radiation and YAG laser pulse.** (a-c) ruby fluorescence. (d) YAG laser pulse. In grayscale of (a-d), white color is brightened in proportion to light intensity. The ratio in (c) is 64 times that in (b), which is 256 times that in (a). t = 0 is the center of the laser pulse.

**3.2.2 Measurements of ruby fluorescence spectrum and generated pressure**

Pulsed YAG laser light with a pulse width of 0.4 ms and $2.1 \times 10^7$ W/cm$^2$ was irradiated through the sapphire anvil to the glassy carbon layer of the sample. Figures 8 (a), (b) and (c) show CCD camera records of fluorescent line spectra from ruby particles at three different grayscales. Figure 8 (d) shows CCD camera record of YAG laser light. Figures 8 (a)-(c) show that ruby fluorescence is emitted over a wide wavelength range simultaneously with the YAG pulse irradiation, the fluorescence intensity drastically decreases once, and the fluorescence with wavelength longer than 800 nm is not observed (see Fig. 8 (c)). Furthermore, approximately 0.5 ms later, the ruby fluorescence is restored again. Figures 8 (a) and (b) show that after the pulse laser irradiation, the intensity of the strong ruby $R_1$ line peak observed at 699 nm before the pulse laser irradiation abruptly decreased and its peak position shifted to the longer wavelength region. The peak position of the $R_1$ fluorescent line at 699 nm indicates that the pressure inside the sample before the YAG pulse irradiation was 13 GPa.

It has been shown that when the temperature of ruby grains exceeds 400 K, fluorescence due to the $^4T_2 \rightarrow {}^4A_2$ process over a wide wavelength range begins to appear by the thermally activated $^2E \rightarrow {}^4T_2 \rightarrow {}^4A_2$ process [32, 33]. Because of this phenomenon, as the temperature rises, the intensity of the $R_1$ fluorescent line peak decreases. At atmospheric pressure, the position where an intensity of the fluorescence band due to the $^4T_2 \rightarrow {}^4A_2$ process becomes maximum exists near the $R_1$ fluorescent line peak. Figures 9(a)-(d) show time dependent change of the ruby fluorescence spectra measured by the CCD. Here, t = 0 ms is the time of the center of the YAG pulse. Figure 9 (a) shows the ruby fluorescence spectrum before the YAG laser irradiation, the ruby fluorescence spectra at the time of t = -0.06, t = 0.03, t = 0.12, and t = 0.24 ms. By comparing the ruby fluorescence spectrum before pulsed YAG laser irradiation and the ruby fluorescence spectrum at the time of t = 0.24 ms (see Fig. 9(a)), it can be found that the amount of increase in fluorescence intensity at the wavelength of 710 nm near the $R_1$ line is about one-third of that of increase in fluorescence intensity at the wavelength of 744 nm. Therefore, the position, at which the intensity of the fluorescence band over a wide range generated after the pulse YAG laser irradiation becomes maximum, exists in the longer wavelength region than 744 nm. Therefore, the fluorescence over a wide range generated after the pulsed YAG laser irradiation is not due to the thermally activated $^4T_2 \rightarrow {}^4A_2$ process. It can be considered that this broad fluorescent band mainly consists of R fluorescent line, and its spread is caused by a wide pressure distribution. In fact, Fig. 9 (b) shows that a peak appears at 905 nm in the ruby fluorescent line spectra at t = 0.21 ms and t = 0.24 ms. If the fluorescence over a wide range is the fluorescence band generated by the thermally activated $^4T_2 \rightarrow {}^4A_2$ process, such peaks or edges cannot appear near the ends. When the above peak is attributable to the $R_1$ line, the generated pressure can be obtained using the relationship between the $R_1$ line peak wavelength and the generated pressure [34], which is used under the non-hydrostatic pressure condition. It is calculated from the peak wavelength of 905 nm by extrapolation, which is 1.05 TPa. It can be estimated that the maximum pressure value of the pressure distribution generated in the ruby layer was 1.05 TPa. The minimum pressure value of the pressure distribution is determined by the peak position around 699 nm (see Fig. 10). The peak at 905 nm is observed in the ruby fluorescence spectra at t=0.21 ms and t=0.24 ms (Fig. 9 (b)). The fact suggests that the maximum pressure of 1.05 TPa was maintained for more than 0.03 ms.

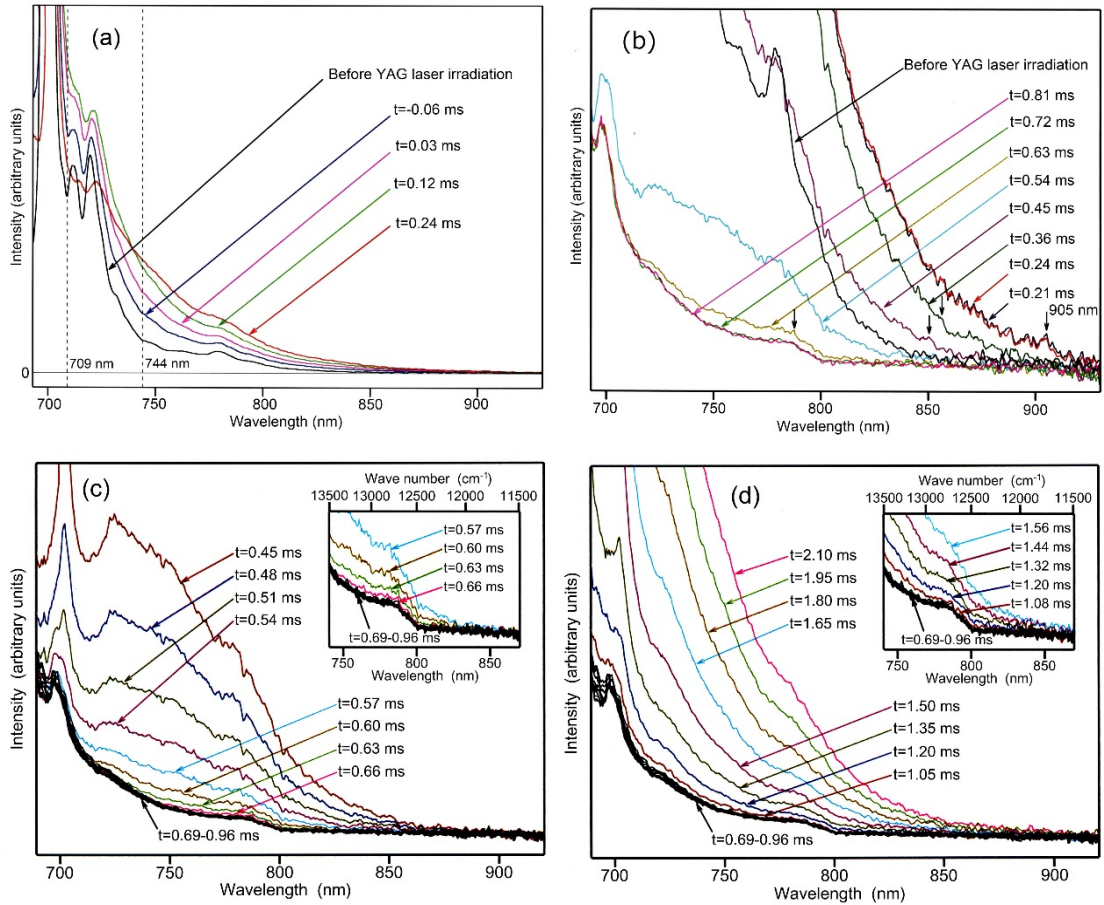

Figure 9 **Changes with time of ruby fluorescence spectra.** (a) t = -0.06 ms ~ t = 0.24 ms, (b) t = 0.21 ms~ t = 0.81 ms, (c) t = 0.45 ms ~ t = 0.96 ms, and (d) t = 0.69 ms ~ t = 2.10 ms. Insets of (c) and (d) show enlarged views around 800 nm. These spectra were corrected with respect to the characteristics of the optical system and the CCD as in the case of the thermal radiation spectrum measurement.

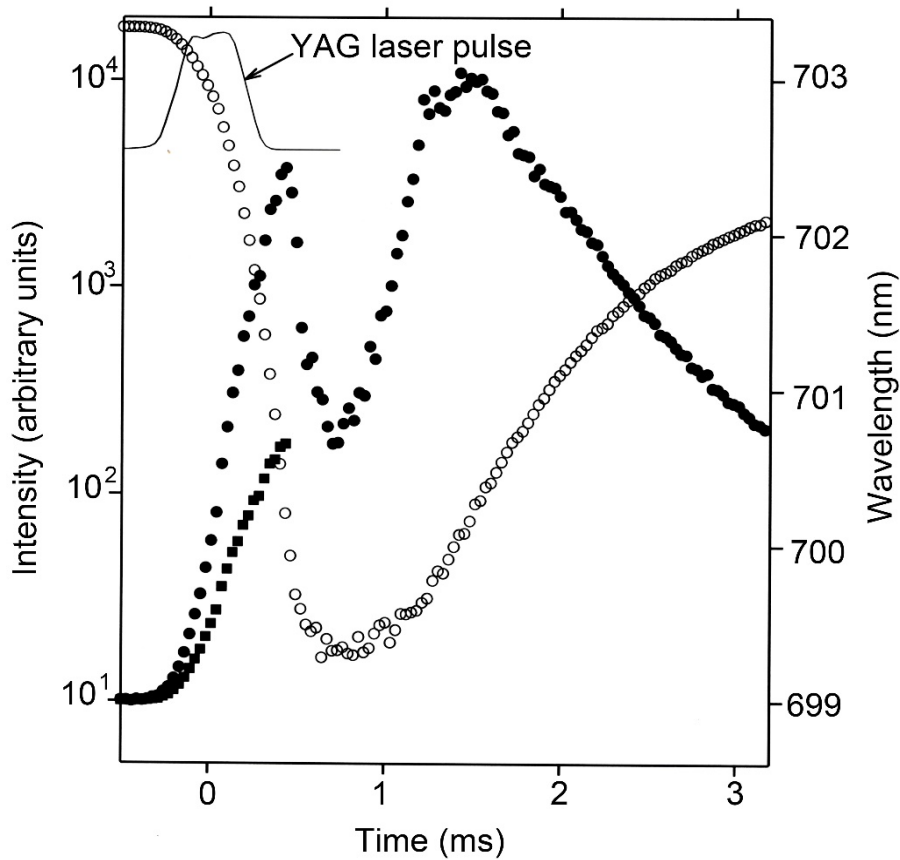

Figure 10 **Time variation of the peak position and intensity of ruby fluorescent $R_1$ line.** ● indicates the peak position of the $R_1$ line, and ○ indicates the peak intensity. The pulse shape of the pulse YAG laser is shown in arbitrary units. t = 0 is the time at the center of the laser pulse. ■ indicates the change with time of only the temperature component of the $R_1$ line shift between t = -0.2 ms and t = 0.48 ms obtained by calculation (see text).

Figure 10 shows the changes with time in the peak position and intensity of the $R_1$ fluorescent line. The peak position of the $R_1$ line was determined by fitting the Gaussian curve to the peak curve of the $R_1$ fluorescent line by using the nonlinear least square method. It is seen from Fig. 10 that simultaneously with the pulse YAG laser irradiation, the peak position of the $R_1$ line started moving toward the long wavelength side and reached the wavelength of 702.434 nm at t = 0.48 ms, after that it sharply shifted to the shorter wavelength side and then moved to a wavelength of 700.657 nm at t=0.75 ms. Furthermore, the peak position of the $R_1$ line shifted again to the longer wavelength side from t = 0.75 ms, the peak wavelength of the $R_1$ line became the maximum of 703.015 nm around t = 1.53 ms, it did not change for a while, and after that the peak position gradually shifted to the shorter wavelength side. The peak position of the $R_1$ line shifts to the longer wavelength side not only by the pressure but also by the temperature. Therefore, it is impossible to simultaneously determine the temperature and the pressure of ruby grains from the peak position of $R_1$ line. The configuration of this experiment is the same as that of the thermal diffusivity measurement system [35] using the laser flash method. In the laser flash method, a pulsed laser is irradiated on the front surface of a thin disk sample, and at the same time, the temperature of the back surface of the disk sample is measured. The $R_1$ line position curve from t = 0.75 ms to t = 1.53 ms and the $R_1$ line position curve after t = 1.53 ms closely resemble the temperature curve of the sample back surface measured by the laser flash method when the sample insulation is incomplete [36] (see Fig. 10). From this fact, it is considered that the shift of the peak position of the $R_1$ line to the long wavelength side after t=0.75 ms is due to the temperature rise caused by the heat transfer from the YAG pulse laser irradiated surface. Furthermore, it can be inferred that the shift of $R_1$ line peak position to the shorter wavelength side after t=1.53 ms is due to temperature decrease caused by heat conduction to the sapphire anvil.

Next, we consider whether a high temperature has occurred at the same time as the generation of an ultrahigh pressure of 1.05 TPa. If we assume that the shift to the shorter wavelength side of the $R_1$ line after t = 0.48 ms is due to a decrease in temperature, it is due to heat conduction to the sapphire anvil, so the slope of the $R_1$ line peak position change curve from t = 0.48 ms to t = 0.75 ms must be almost equal to that of the curve after t = 1.53 ms. However, the slope around t = 0.48 m of the $R_1$ line peak position change curve from t = 0.48 ms to t = 0.75 ms is about 7 times that of the $R_1$ line peak position change curve after t = 1.53 ms. From this fact, it is considered that the change of the $R_1$ line peak position from t=0.48 ms to t=0.75 ms is not due to a temperature drop but due to a decrease in pressure. Therefore, the change amount of the $R_1$ line peak position between t=0.48 ms and t=0.75 ms can be regarded as the pressure component of $R_1$ line shift at t=0.48 ms. The value obtained by subtracting the difference of the $R_1$ line peak wavelength from t=0.48 ms to t=0.75 ms from the $R_1$ line peak wavelength at t=0.48 ms can be considered to be approximately equal to the temperature component of the $R_1$ line peak shift at t=0.48 ms. We assume that the ratio of the pressure component and the temperature component of the $R_1$ line shift amount between the shift start time of $R_1$ line after pulsed YAG laser irradiation and time t = 0.48 ms is constant. Furthermore, we assume that the above ratio is equal to the ratio of the pressure component and the temperature component of the $R_1$ line shift amount at t=0.48 ms. Based on these assumptions, the $R_1$ line peak position due to only the temperature shift at each time from

t=-0.15 ms to t=0.48 ms was obtained and these results are plotted by ■ in Fig. 10. In Fig. 10, it seems that the plots indicated by ■ up to t=0.48 ms and the plots indicated by ● from t=0.75 ms do not continuously connect. This is because the assumed pressure component of the $R_1$ line shift at t = 0.48 ms (i.e., the change amount of the $R_1$ line peak wavelength between t = 0.48 ms and t = 0.75 ms) was smaller than the actual pressure component at t=0.48 ms. From the above discussion, it can be inferred that the expansion force due to the decomposition of glassy carbon by pulse laser irradiation first propagates to the ruby particle layer and the effect of heat due to thermal conduction appears on the ruby particle layer later than the influence of the expansion force. Therefore, the $R_1$ line peak shift at t=0.48 ms is almost due to pressure increase, not due to temperature rise. Since the temperature of the sample rose almost smoothly after the pulse laser irradiation (see the plots ■ in Fig. 10), it can be judged that there is no occurrence of high temperature due to the influence of heat caused by generation of ultrahigh pressure (1.05 TPa) at t = 0.21 ms or t = 0.24 ms.

Figure 9 (c) shows how the longest wavelength position of the $R_1$ fluorescent line after t = 0.24 ms gradually moves toward the shorter wavelength side. After t = 0.45 ms, an edge with a steep slope appears near 800 nm in the long wavelength region, indicating that fluorescence having a wavelength of about 800 nm is continuously emitted. Since there is small offset between the $^4A_2$ and $^2E$ states, the $^2E \rightarrow ^4A_2$ emission (R line) is a sharp line. Therefore, it is considered that the steep slope near 800 nm is characteristic of the sharp R line peak (corresponding $^2E \rightarrow ^4A_2$ transition). The pressure in the ruby layer is thought to be distributed widely and continuously, and this edge corresponds to the slope on the longer wavelength side of the $R_1$ line peak emitted from the ruby which is pressured to the maximum pressure after t = 0.45 ms. Half of the full width at half maximum of the peak at the edge is obtained from this edge curve. Compared with this value and half of the full width at half maximum of the R line peak observed at 220 GPa [37], it is found that the values of both are very close. Based on the ruby $R_1$ fluorescent line scale of non-hydrostatic pressure [34], the edge wavelength 785 nm (see inset in Fig. 9 (c)) corresponds to 323 GPa, which shows that the maximum pressure of 323 GPa was generated. The higher the pressure, the longer the fluorescence lifetime. Furthermore, the fluorescence lifetime of the $R_1$ line is ~ 3 ms at atmospheric pressure [38]. The emission of R fluorescent light from the high-pressure region of 323 GPa is impossible to excite with excitation light of 514.5 nm, but the emission occurs by the transition from $^2E$ to $^4A_2$ state because $^2E$ state is occupied by $^4T_2 \rightarrow ^2E$ transitions before the pulse laser irradiation. Figure 9 (d) shows how the edge disappears due to pressure attenuation after t = 1 ms. It is seen that the edge was observed from t = ~ 0.5 ms to ~ 1.5 ms and the pressure of 323

GPa was sustained for about 1 ms (see Figs. 9 (c) and (d)).

It has been pointed out that sapphire becomes opaque from the pressure of 100-130 GPa due to internal defects caused by impact stress [39]. Therefore, since no fluorescence was observed for about 0.5 ms, it might be assumed that sapphire anvil became opaque due to the generation of pressure exceeding 100 GPa. The temperature rises at times between t = 0.75 ms and t = 1.53 ms (see Fig. 10). However, although the strength of the $R_1$ line should decrease due to the temperature rise, the intensity of the $R_1$ line increases during the above-mentioned time. This means that the transparency of sapphire has recovered by the decrease in pressure. However, if the cause of opaqueness of the sapphire anvil is due to defects, the defects once formed cannot disappear, so it is impossible to explain that the transparency has recovered, and ruby fluorescence becomes observable. While ruby fluorescence with a range of wavelengths could not be observed, the maximum pressure of the ruby layer was maintained at 323 GPa. The pressure of 323 GPa is close to the metallization pressure which is theoretically predicted for alumina single crystal (sapphire) [40, 41]. Therefore, it may be presumed that the vicinity of the pressure plane of sapphire anvil became opaque due to narrowing of the band gap of sapphire caused by high pressure of 323 GPa.

**3.3 Summary**
(1) From the measurement of the radiation spectrum, it was found that sufficient high temperature was generated to decompose interatomic bonds in the carbon layer by pulse laser irradiation.
(2) Ruby fluorescence measurement showed that the maximum pressure of 1.05 TPa was maintained for 0.03 ms or more when sapphire anvil cell was used for constraining the decomposition of the glassy carbon pressure source.
(3) There is almost no heating accompanying the generation of the above pressure. The temperature rise of the pressurized specimen is mostly due to heat conduction from the pulsed laser irradiation surface.
(4) The maximum value of the pressure distribution quickly decreased to 325 GPa after rising to 1 TPa, and this pressure was maintained for 1 ms.
(5) While the pressure of 325 GPa is being generated, the vicinity of the sapphire anvil pressure plane is considered to be opaque to long wavelength light.

**4. Metallization experiment of diamond**
**4.1 Experiment**
Figure 11 shows a method for measuring the change in reflectance of a diamond single crystal placed in a sample chamber which is caused by irradiating a glassy carbon layer with a pulsed YAG laser. In this experiment, diamond fine powder (average particle size: 1 μm) was selected as the pressure medium to pressurize the diamond single crystal, and sapphire and diamond anvils were used as the high-pressure generation anvil. Here, the diamond single crystal was placed on the side of the diamond anvil and its crystal face was brought into close contact with the pressure face of the diamond anvil. Therefore, the crystal plane and the pressure plane are parallel to each other. For reflectance measurements, the beam from a He-Ne laser (wavelength: 632.8 nm) was focused on the diamond single crystal through the diamond anvil so that its direction was perpendicular to the diamond anvil pressure face. Light, which is reflected from the diamond single crystal, was detected by CCD using the same optical system as ruby fluorescence measurement.

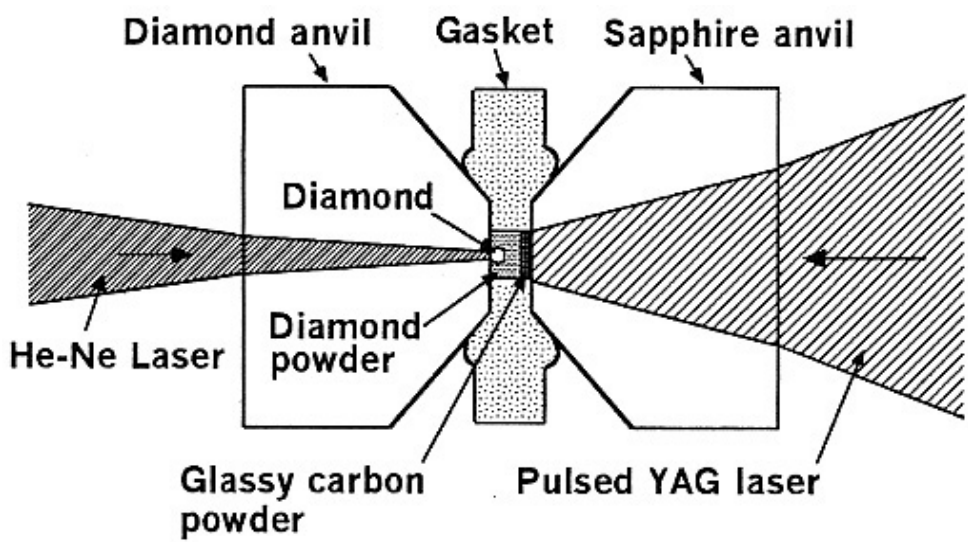

Figure 11 **Method for measuring reflectance from single crystal diamond placed in sample chamber.** The (100) surface of the diamond single crystal is in direct contact with the diamond anvil pressure surface in parallel.

**4.2 Results and discussion**
**4.2.1 Changes in reflectance of diamond due to high pressure generation**

In order to pack the diamond fine powder at high density into the pressure chamber, a pulsed YAG laser light of pulse width 0.4 ms, $2.1 \times 10^7$ W was first irradiated through the sapphire anvil to the glassy carbon layer. Figure 12 shows the change with time in the reflectance from the boundary between the diamond crystal and the diamond anvil after the second pulse laser irradiation. In order to determine the reflectance at the interface between the diamond single crystal and the diamond anvil pressure surface, it was assumed that the refractive index of the diamond anvil was constant at 2.42 and the refractive index of the atmosphere was 1 during high pressure generation. The reflectance was determined from the intensity of the light emitted from the diamond anvil as follows.

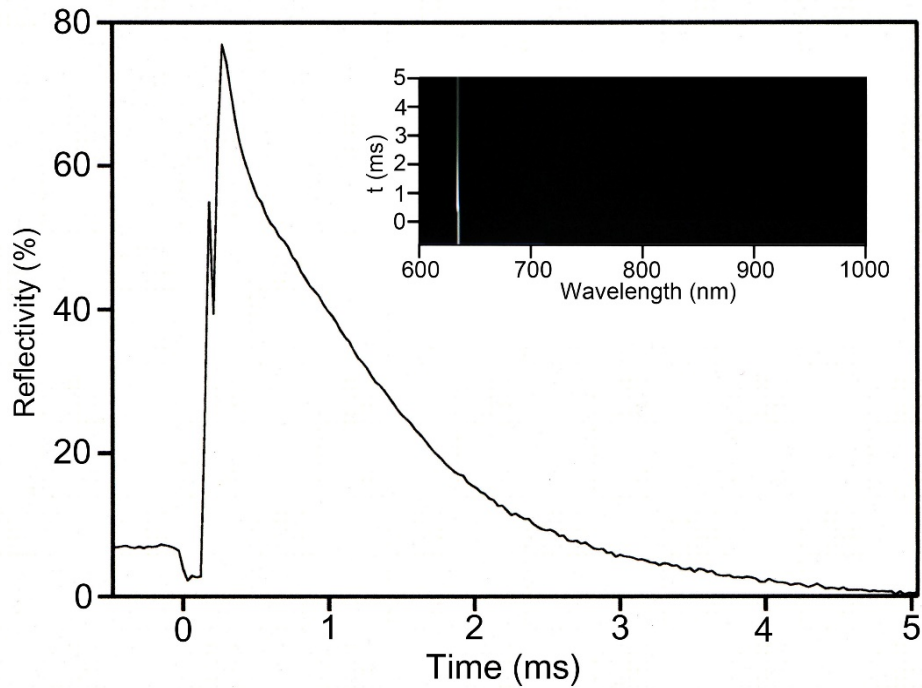

Figure 12 **Change with time of reflectance at the boundary between diamond single crystal and diamond anvil pressure surface after pulse YAG laser irradiation.** The inset shows the CCD recording of reflected light of He-Ne laser light, and the change with time was obtained based on this CCD data.

Since ultrahigh pressure is not generated before the pulse laser irradiation, all the incident He-Ne laser beams that transmitted through the diamond anvil passes through the boundary between the diamond single crystal and the diamond anvil pressure surface. Therefore, it is considered that the light emitted from the back side of the diamond anvil is only the light reflected from the anvil's back surface. The reflectance r on the back surface of the diamond anvil is expressed by the following equation (7) using the refractive index n of the diamond.

$$r=\{(1-n)/(1+n)\}^2 \quad (7)$$

When the intensity of the irradiated He-Ne laser beam is represented by $I_0$, the reflected light intensity I' from the back surface of the diamond anvil before the pulse laser irradiation is given by the following equation (8).

$$I'=rI_0 \quad (8)$$

The irradiated He-Ne laser light passes through the diamond anvil and reaches the boundary surface between the diamond single crystal and the diamond anvil. The intensity I″ of the light that reaches the interface is given by the following equation (9).

$$I'' = (1-r)I_0 \qquad (9)$$

If the reflectance at the boundary surface between the diamond single crystal and the diamond anvil pressure surface after the pulse laser irradiation is R, the intensity of the reflected light from this boundary surface is expressed by $(1-r)RI_0$. When this reflected light passes through the boundary between the back surface of the diamond anvil and the atmosphere and exits from the diamond anvil into the atmosphere, its intensity $I_1$ is given by the following equation (10). At the same time, the intensity $I_1′$ of the light reflected by the boundary between the back surface of the diamond anvil and the atmosphere can be obtained by the following equation (11).

$$I_1 = (1-r)^2 R I_0 \qquad (10)$$
$$I_1' = (1-r)rRI_0 \qquad (11)$$

Furthermore, the light reflected at the boundary between the back surface of the diamond anvil and the atmosphere again reaches the interface between the diamond single crystal and the diamond anvil pressure surface, and the intensity of the light reflected by the above interface becomes $(1-r)rR^2I_0$. When this reflected light passes through the boundary between the back surface of the diamond anvil and the atmosphere and comes out into the atmosphere from the diamond anvil, its intensity $I_2$ is given by the following equation (12), and the intensity of the light $I_2'$ that is reflected by the boundary between the back surface of the diamond anvil and the atmosphere is obtained by the following equation (13).

$$I_2 = (1-r)^2 r R^2 I_0 \qquad (12)$$
$$I_2' = (1-r)r^2 R^2 I_0 \qquad (13)$$

Eventually, the reflection at the interface between the diamond anvil's back surface and the atmosphere and at the interface between the diamond anvil's pressure surface and the diamond single crystal is repeated, and the intensity I‴ of the reflected light from the boundary surface between the diamond anvil and the diamond single crystal that exits into the atmosphere from the diamond anvil is given by the following equation (14).

$$I''' = (1-r)^2 R I_0 (1 + rR + r^2 R^2 + \cdots r^n R^n + \cdots) = (1-r)^2 R I_0 / (1-rR) \qquad (14)$$

Furthermore, since $0 < rR < 1$, the intensity I‴ of the reflected light is obtained by the following equation (15).

$$I''' = (1-r)^2 R I_0 / (1-rR) \qquad (15)$$

If the intensity of the reflected light from the back surface of the diamond anvil is added this intensity (I‴), the intensity I of the light emitted from the back surface of the diamond anvil after the pulse laser irradiation is given by the following equation (16). Here, I´ is the intensity of reflected light from the back surface of the diamond anvil before the pulse laser irradiation.

$$I=rI_0+(1-r)^2RI_0/(1-rR)= I´+(1-r)^2R\ I´/\{r(1-rR)\} \quad (16)$$

From the above equation, the reflectance R at the interface between the diamond single crystal and the diamond anvil pressure surface after the pulse laser irradiation is given by the following equation (17) when α=I/I'.

$$R=r(\alpha-1)/(1-2r+\alpha r^2) \quad (17)$$

The inset of Fig. 12 is a CCD recording of reflected He-Ne laser light, which shows CCD recording in the wavelength range from 600 nm to 1000 nm. The reflectance suddenly increases at t = 0.09 ms, reaches a maximum value of 76% at about 0.15 ms, and after that gradually decreases to zero. This high reflectance is considered to suggest metallization of diamond. From the inset of Fig. 12, it can also be seen that thermal radiation from the backside of the carbon layer heated to 9000 K was not observed through the diamond fine powder. The reason why very strong thermal radiation was not observed is probably because the single diamond crystal and diamond fine powders had high light absorption or high reflectance due to the addition of uniaxial stress due to the expansion force generated by decomposition of the glassy carbon layer and they shielded the thermal radiation. The fact that this thermal radiation was not observed and the result of the reflectance measurement of the diamond single crystal can be evidence that high pressure causing metallization of diamond has been generated in the pressure chamber.

The metal transition pressure of diamond was estimated to be 700-900 GPa from experimental extrapolation [42]. On the other hand, the band gap of diamond is predicted to increase by pressure increase even in the ultrahigh pressure region exceeding 1 TPa [43]. However, it is shown that when the diamond is subjected to uniaxial pressing, symmetry of the diamond changes and the sign of the rate of change with respect to the pressure of the band gap changes to negative [44]. It is theoretically predicted by Nielsen that metallization of diamond occurs even with uniaxial stress of 400 GPa [45]. In the experiments presented, it seems that isotropic pressure and uniaxial stress were instantaneously applied to the diamond single crystal and the diamond fine powder surrounding it due to the expansion force of the carbon layer caused by the pulsed laser irradiation. In this experiment, metallization of diamond may have occurred due to uniaxial stress of 400 GPa or more applied to the diamond single crystal.

From the inset of Fig. 12, the thermal radiation from the diamond single crystal was not recorded in the CCD up to the wavelength of 1000 nm. Based on the blackbody radiation curve, the CCD image suggests that the temperature of the diamond single crystal did not exceed 700 K during the experiment. Since the temperature of the single crystal before the YAG laser irradiation was about 300 K, the rise in temperature of the single crystal due to heat accompanying the generation of high pressure is considered to be lower than 400 K. This rise in temperature is much lower than the rise in temperature accompanying the generation of pressure in the conventional dynamic compression method.

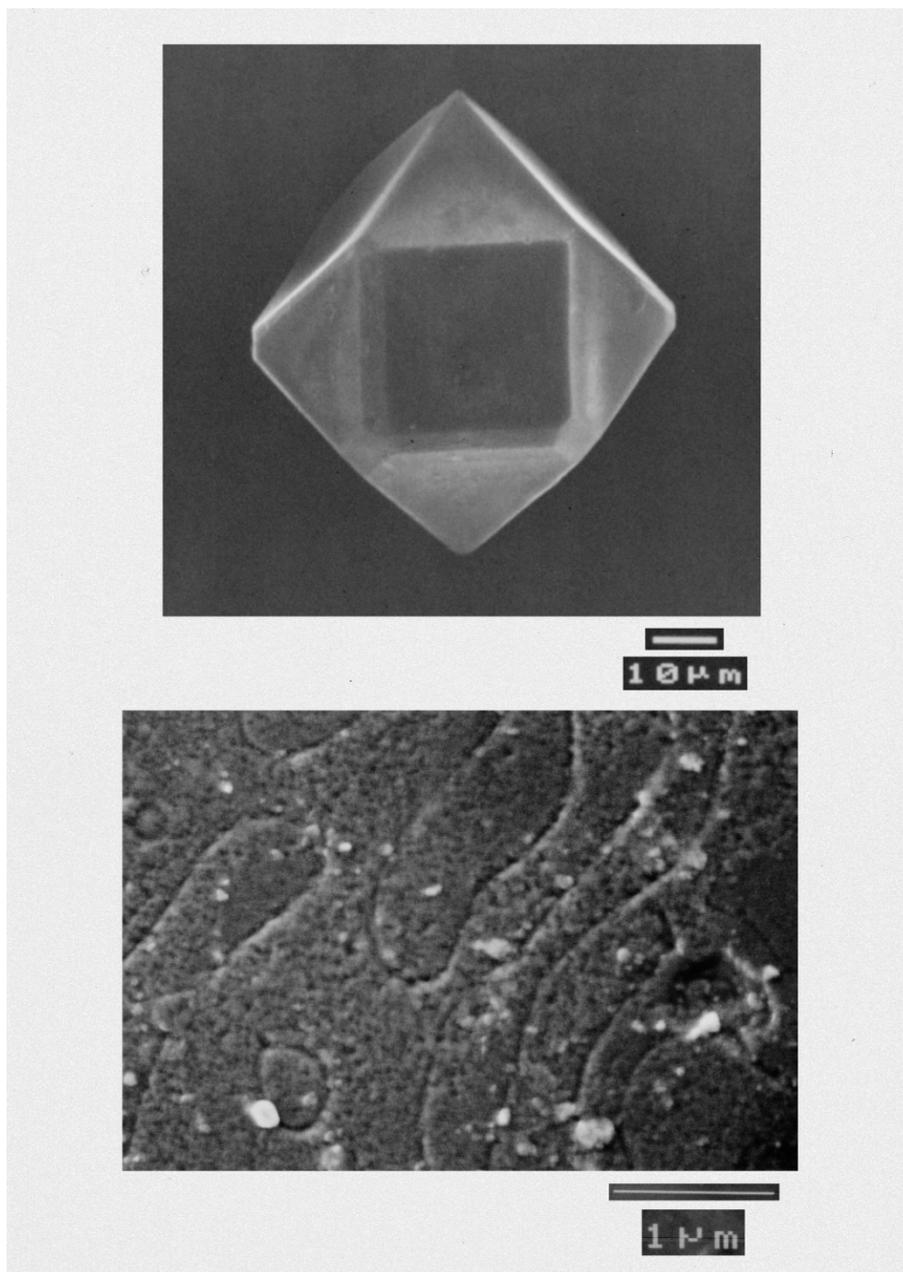

Figure 13 **SEM photograph of diamond single crystal after being subjected to ultrahigh pressure and the enlarged SEM photograph of its (100) surface.**

## 4.2.2 Dislocation density of diamond single crystal to which ultrahigh pressure was applied

The diamond single crystal after exposure to high pressure by this high-pressure apparatus did not change the transparency but had a very large dislocation density comparable to the plastically deformed metal. Figure 13 (a) is a SEM photograph of a diamond single crystal taken after being exposed to high pressure, and Fig. 13 (b) shows a SEM photograph taken at a higher magnification of its (100) surface. Dislocation loops and etch pits are observed on the (100) surface of the diamond single crystal exposed to high pressure (see Fig. 13 (b)). The dislocation density is estimated from the number of etch pits to be $10^{11}$ cm$^{-2}$. This value is comparable to that of heavily deformed metals [46]. This result shows that the diamond single crystal was metallized and subsequently experienced a large deforming force, which means that it was plastically deformed in the metallic state. The trace of the large plastic deformation observed shows strong evidence of metallization of the diamond single crystal.

## 4.3 Summary

The high-pressure experiment presented in this chapter showed that the reflectance at the interface between the diamond anvil pressure surface and the diamond single crystal rises to a value comparable to the reflectance of the metal. Dislocations indicating that very large plastic deformation occurred in the pressed diamond single crystal were observed. The high reflectance and the large dislocation density observed for the pressed diamond crystal can be explained only by metallization of diamond. Therefore, from the metallization of diamond, it can be concluded that uniaxial stress of 400 GPa or more has occurred.

## 5. Conclusion

Aiming for the synthesis of solid metal hydrogen, we introduced a high-temperature and high-pressure apparatus that vaporizes solids in a constrained state, decomposes them into atomic states separated into pieces, and constrains the explosion force produced thereby. This apparatus can generate a high pressure of 1 TPa or more and has a high-pressure generation capability far higher than that of the diamond anvil cell device. Furthermore, by using a sapphire anvil cell as a device for constraining expansion, it is possible to secure a high-pressure chamber having a volume several orders of magnitude larger than that of a diamond anvil cell device. Although this apparatus cannot statically generate pressure unlike the diamond anvil device, it can maintain the high pressure for several orders of magnitude longer than the conventional dynamic compression apparatus.

Furthermore, unlike the conventional dynamic compression apparatus, in this apparatus the generation of high pressure is not accompanied by the generation of high temperature. If the anvil cell of the present apparatus is put in a cooling device and the experiment is carried out, a high pressure of 1 TPa can be maintained for 0.03 ms or more at a temperature of room temperature or lower. If the theoretical prediction that the synthesized solid metal hydrogen is metastable even after removing pressure is correct, it can be deduced that solid metal hydrogen might be synthesized by maintaining the solid metal hydrogen state even for a very short time. Therefore, since the present apparatus can generate high pressure of 1 TPa for 0.03 ms or more at at a temperature of room temperature or lower, it may be able to synthesize solid metal hydrogen. In addition, relatively large volumes of solid metal hydrogen might be obtained by using a large sapphire anvil. Thus, this apparatus can generate high pressure necessary for solid hydrogen synthesis which cannot be achieved by the diamond anvil cell apparatus and furthermore it can avoid the generation of high temperature accompanying generation of high pressure which is unavoidable in the conventional dynamic compression apparatus. In this paper, we introduced a new type of high-temperature and high-pressure apparatus that may enable solid metallic hydrogen synthesis.

**Acknowledgments**
This research was partially supported by the Ministry of Education, Science, Sports and Culture of Japan, Grant-in-Aid for Scientific Research (A), 1994-1996 (No. 06559013, Yasushi Kawashima), Grant-in-Aid for Scientific Research (B), 1996-1997 (No. 08459023, Yasushi Kawashima), Grant-in-Aid for Scientific Research (B), 1999-2001 (No. 11490032, Yasushi Kawashima), and Grant-in-Aid for Scientific Research (A), 2002-2004 (No. 14209014, Yasushi Kawashima). I would like to thank R. Shimidzu of PHOTON Design Corporation for help with optical design.